# Mechanical properties of metal decorated graphyne, graphyne-BN-yne, and BN-yne sheets

Aidin Ahmadi [a], Hamid Ghorbani Shiraz [b], Mahdi Faghih Nasiri [c]*, Moones Sabeti [d]

[a] Young Researchers and Elite club, South Tehran Branch, Islamic Azad University, Tehran, Iran
[b] Young Researchers and Elite Club, Mashhad Branch, Islamic Azad University, Mashhad, Iran
[c] Young Researchers and Elite club, Central Tehran Branch, Islamic Azad University, Tehran, Iran
[d] Department of Physics, NRTC, Iran

## Abstract

In this paper, the mechanical properties of two-dimensional structures of metal decoration systems of simple graphyne (CC), analogous system of BN sheet (BN-yne), also the graphyne-BN sheet (CC-BN-yne) was investigated. The properties such as Young and Bulk moduli were studied using Energy-Strain correlation. We introduced calculations based on density functional theory (DFT); the generalized gradient approximation (GGA) framework was used in this regard. The results demonstrated very competitive values for Young and Bulk moduli of the Pt decorated CC and BN-yne. However, the CC-BN-yne structure defined around 80% of Young and 77% of Bulk values of that of pure structures. Also Na decorated ones were examined and the results showed the same trend for all three structures. The CC-BN-yne defined the lowest values for either Young or Bulk moduli.

## Introduction

Todays, wide range of desirable properties has defined carbon nanostructures (CNSs) as one of the high usable materials in several industries such as solid state physics and materials science [1]. Like other nanomaterials, CNSs are divided in four distinct categories of 0D (Fullerene), 1D (Carbon nanotube), 2D (Graphene) and 3D (Diamond, Graphite) [2]. In the last years, two dimension CNSs such as graphene have received high amount of attentions. Several properties like high charge mobility, desirable mechanical flexibility, and high surface area [3] have made graphene as a potential material for the applications such as high-sensitive mechanical sensors [4, 5] such as robots [6, 7], motion sensors [8-11], and seismographic systems [12]. Elastic characters of nanostructures could be measured by considering of interfere forces and correlating the (sheer) stresses and relative length of the structure. One of the most important criterions that may introduce the elasticity properties is the Young's modulus. In case of graphene, Young's modulus is defined using experimental methods and AFM (Atomic Force Microscopy). Graphyne could be assigned as a new member of planar carbon nanostructure. It has been recently fabricated by chemical vapor deposition method and simulated hot researches in theoretical and practical [13, 14]. Graphyne is very similar to the graphene, structurally. Graphene include sp2 carbon atoms hybridization while graphyne consist of both sp and sp2 hybridized carbon atoms. These rate of resemble have proposed high degree of similarity in properties such as mobility, stability, and mechanical properties. However, the presence of acetylenic bonds define numerous physical properties [15-20]; what are different from that of have been found in graphene.

Puigdollers et al. [21] studied different properties of structural, mechanical, and electronic of the graphyne using Density functional theory (DFT); also, they compared the results with those of graphene. It was demonstrated that graphene might offer superior value for in-plane stiffness. The inclusion of acetylinic linkages in graphene lattice decreases the number of bonds and the planar density and reduces the rigidity of the material. Also, the Fermi velocities found higher in the case of graphene. In fact, graphyne is a semiconductor with the band gap of 0.46 eV in the M point. This fact, as well as the anisotropy of the carrier makes graphyne a very interesting material due to electronic properties. While the graphyne presented higher Poisson's ratio of 0.87, compared to that of the graphene. This phenomenon was ascribed to the sparsity of in-plane atoms in graphyne that permits larger contractions. Also, Asadpour et al. [22] study the mechanical properties for graphyne, graphyne-like BN sheet and analogous system of BN sheet, based on the DFT calculations. They calculated in-plane stiffness through variation of strain energy and demonstrated that graphyne reveals highest in-plane stiffness with the value of 190.69 N/m. in addition, largest poisson's ratio and modulus (both sheer and bulk) obtained through Analogous system of BN sheet and graphyne, respectively.

---

*Corresponding Author's emails: mahdi.faghihnasiri@gmail.com
mahdi.faghihnasiri@shahroodut.ac.ir

Recently, decorated systems have been introduced in computational studies. By the referring to the literature, several studies could be found in this regard. In fact, the variation of intrinsic properties could be such a way that desirable features could be provided, and this is attributed to the decorated species. Recently, it has been proved that Na-decorated graphyne can be employed as carbonaceous compound conductor. This has been attributed to the charge donation from sodium to carbon [23]. Also, the adsorption of hydrosolforic over graphene system has been investigated, with the aid of DFT calculations [24]. The scientists proved that orbital hybridization could be performed between $H_2S$ and Pt-decorated graphene, while such stable hybridization could not be occurred between $H_2S$ and non-decorative sample. As far as we know, there is no report for consideration of mechanical properties for Na- and Pt-decorated graphyne and analogous systems.

In this study, we consider computational results of electronic and structural properties of metal decorated 2D-graphyne and its analogous structures involving BN rings (BN-yne). The third structure is introduced as BN hexagonal configurations jointed by C-chains. We suppose that BN sheet as well as its hybrid sheets with graphene have been successfully synthesized [25-27]. Here, the DFT calculations are employed to compute total strain energies for prediction of mechanical properties based on the Perdew–Burke–Ernzerhof exchange correlation.

# Calculation Details and Results

Here, we carried out DFT calculations using Local Combination of Atomic Orbital (LCAO) using the ab initio simulation code, Spanish Initiative Electronic Structure for Thousands of Atoms (SIESTA) [28]. The calculations were performed within the generalized gradient approximation (GGA) as described by Perdew–Burke–Ernzerhof (PBE) [29,30] to study the effects of electronic exchange and correlation.

The split-valence double-ζ basis set of atomic orbitals was introduced, including polarization functions with an energy shift of 50 m eV and a split norm of 0.25 [31]. A 25×25×1 Monkhorst–Pack grid for k-point mesh of the Brillouin zone was proposed, and the local relaxations of atoms were delayed until the residual forces over individual atom reached lower than 0.005 eV/Å. Also, the cut-off energy of 325 Ry for the grid integration was examined in the calculations to represent the charge density. The interlayer vacuum of about 15 Å was introduced to minimize the interactions.

Three aforementioned structures that are shown in the figures 1 to 5 were trade off, and the optimum values for lattice vectors obtained that demonstrate a good agreement with the literatures [32]. Moreover, with the help of calculations, bond lengths obtained in such values that the stable configuration of systems consisted of carbon, Nitrogen, and Boron is recognized. The results for computed lattice vectors and bond lengths have stood in the table 1.

In this investigation, we studied the mechanical properties for Na and Pt decorated systems of CC, BN-yne, and CC-BN-yne. Liu et al. [33] examined different adsorption sites of Na over graphyne, including hexagonal hollow site, triangle hollow site, and the site above acetylenic bond. They demonstrated that favorable site of the spontaneous process of Na adsorption, like other alkali metal adsorption on graphyne [34-37], is the site above the hollow site of triangle formed with acetylenic bonds; because it contributed to the lowest adsorption energy of -2.37 eV.

Thus, we employed this approach in the computation with 2 decorated; below and above nanosheet.

# Mechanical Properties

Mechanical properties of proposed structures were calculated under strain range of [-2% ~ 2%]; so-called harmonic elastic deformation range. Configuration of the structures is kept in the range; consequently, the mechanical properties are stable.

# Conclusion

In this paper, mechanical properties of metal decorated structures of CC, BN-yne, and CC-BN-yne were investigated based on DFT. Variation of strain energy with uniaxial strain resulted in the values of in-plane stiffness. The values for Na and Pt decorated of CC, BN-yne, and CC-BN-yne have stood in table 2. The achieved results showed that the largest values were recognized by CC; higher than that of obtained via non-decorative system. The value of Bulk modulus of graphyne is larger than other structures, when the Na decorated systems is considered. While the Bulk modulus for Pt decorated systems did not defined the same as Na one. This discrepancy may bold the effects of size of decorated metal as well as the electron resonance.

# Figures

**Na decorated structures**

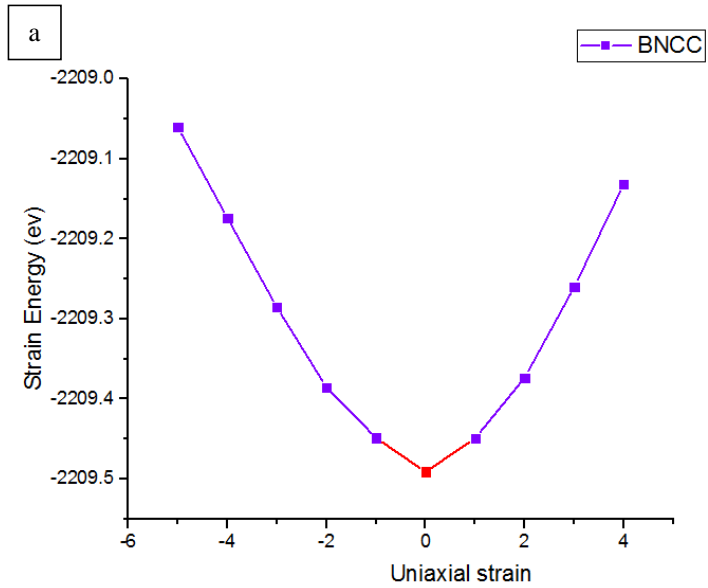

**Figure 6-**a) variation of dE$_s$ in terms of input uniaxial strain for BNCC at the region of homogeneity

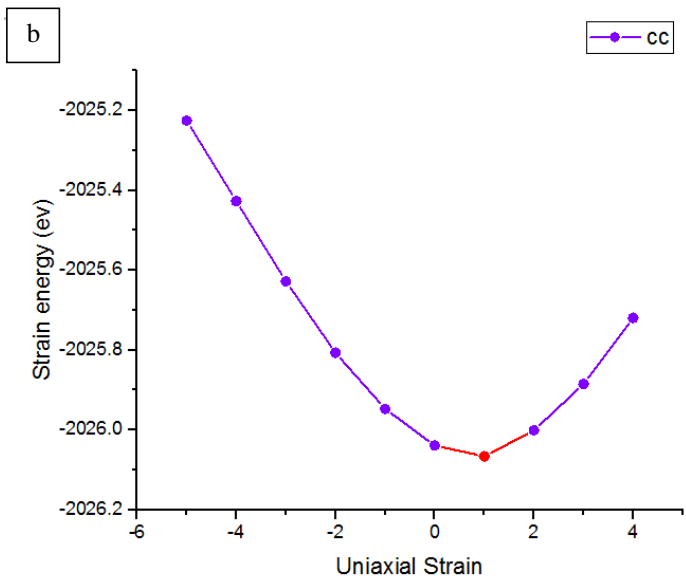

**Figure 6-**b) variation of dE$_s$ in terms of input uniaxial strain for CC at the region of homogeneity

**Pt decorated Structures**

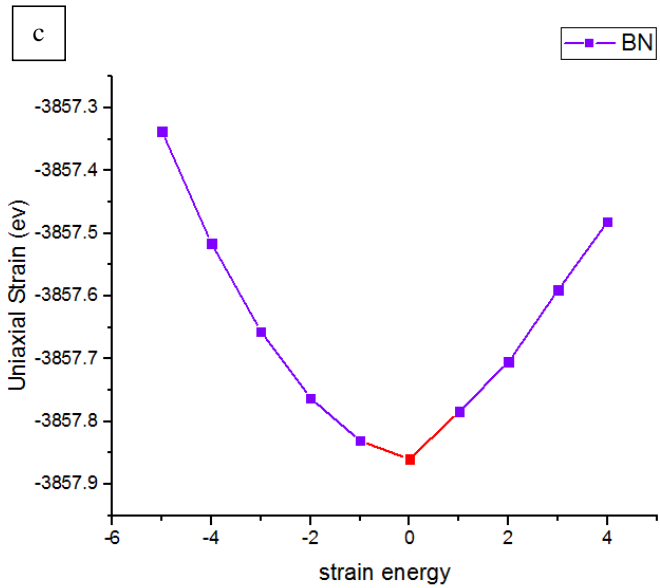

**Figure 6-**c) variation of dE$_s$ in terms of input uniaxial strain for BN at the region of homogeneity

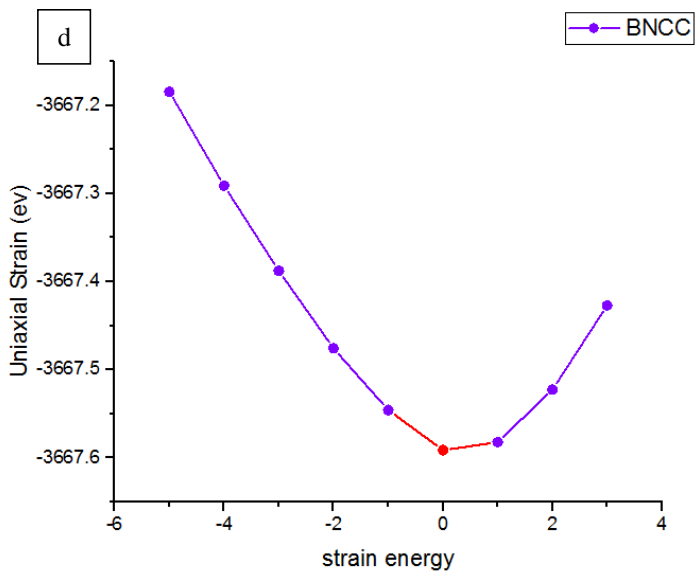

**Figure 6-**d) variation of dE$_s$ in terms of input uniaxial strain for CC-BN-yne at the region of homogeneity

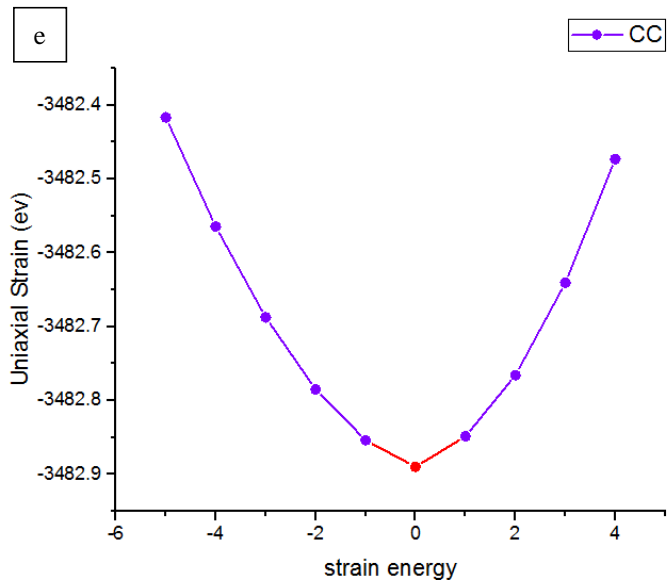

**Figure 6-**e) variation of $dE_s$ in terms of input uniaxial strain for CC at the region of homogeneity

**Na decorated structures**

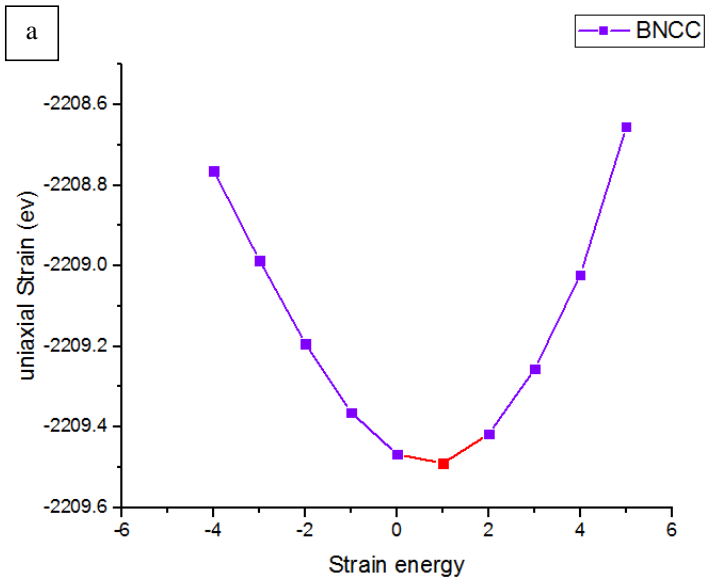

**Figure 12-**a) variation of $dE_s$ in terms of input triple axial strain for BNCC at the region of homogeneity

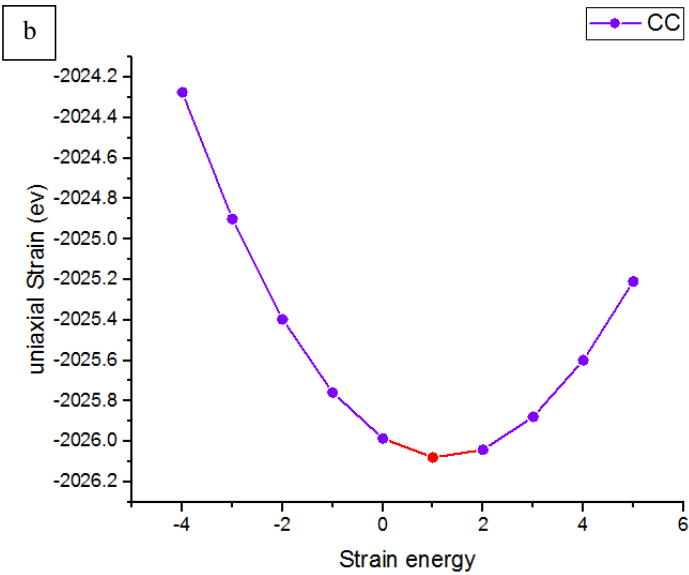

**Figure 12-**b) variation of $dE_s$ in terms of input triple axial strain for CC at the region of homogeneity

## Pt decorated structures

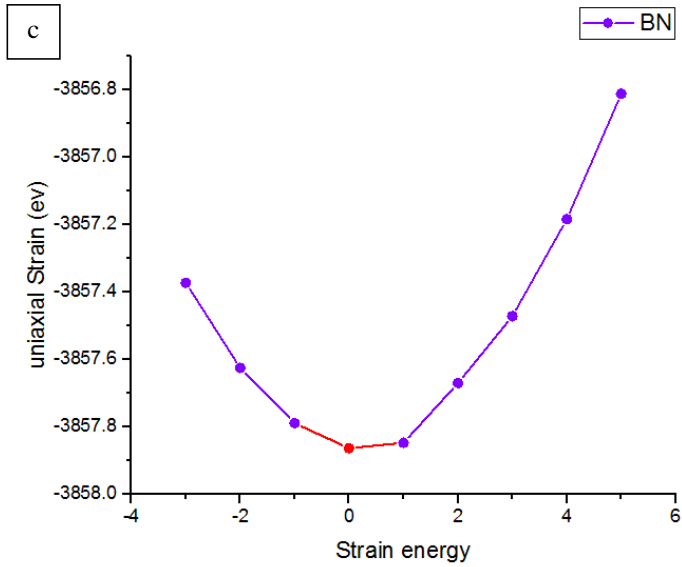

**Figure 12-**c) variation of $dE_s$ in terms of input triple axial strain for BN at the region of homogeneity

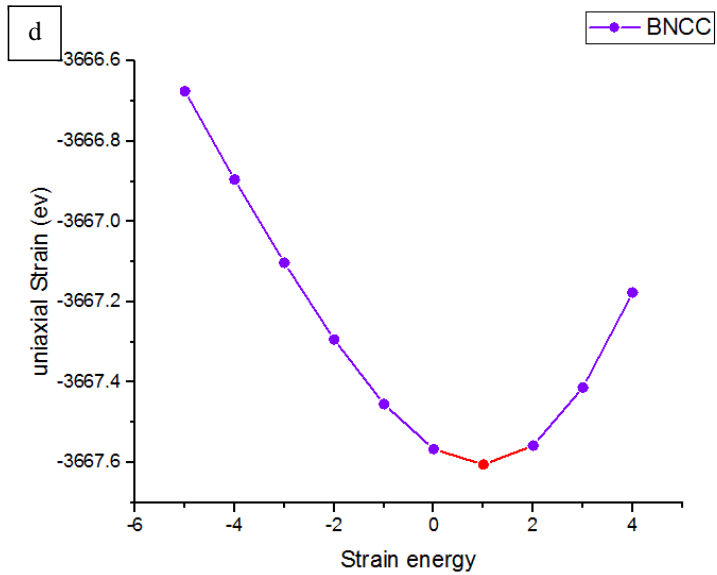

**Figure 12-**d) variation of $dE_s$ in terms of input triple axial strain for CC-BN-yne at the region of homogeneity

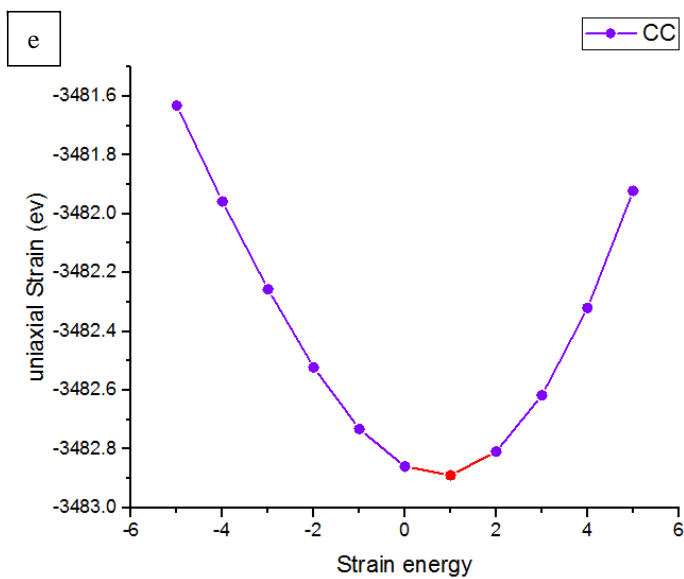

**Figure 12-**e) variation of dE$_s$ in terms of input triple axial strain for CC at the region of homogeneity

*Table 1-Bond length and Lattice vectors for three proposed structure*

| CC | | | |
|---|---|---|---|
| Bond length (Å) | | Lattice Vectors (Å) | |
| C-C | 1.4 | A | 3.43550000 |
| C-C | 1.4 | B | -5.95046000 |
| C-C | 1.4 | C | 20 |

| BN-yne | | | |
|---|---|---|---|
| Bond length (Å) | | Lattice Vectors (Å) | |
| B-N | 1.4 | A | 3.50227000 |
| N-N | 2.67 | B | -6.06611000 |
| B-B | 2.67 | C | 20 |

| CC-BN-yne | | | |
|---|---|---|---|
| Bond length (Å) | | Lattice Vectors (Å) | |
| C-C | 1.24 | A | 3.48948000 |
| C-B | 1.49 | B | -6.04396000 |
| B-N | 1.4 | C | 20 |

*Table 2-In-plane stiffness and Bulk modulus*

|  | Na | | | Pt | | |
|---|---|---|---|---|---|---|
|  | CC | BN-yne | CC-BN-yne | CC | BN-yne | CC-BN-yne |
| In-plane stiffness | 198.42 | ? | 148.09 | 160.39 | 159.20 | 128.54 |
| Bulk module | 109.61 | ? | 68.95 | 79.36 | 80.16 | 61.65 |